\documentclass[12pt]{article}

\usepackage{graphicx}
\usepackage{fancyhdr}   
\usepackage{amsmath}    
\usepackage{comment}    
\usepackage{caption}
\usepackage{fancybox,shadow}
\usepackage{morefloats}

\newcommand{\veps}{\varepsilon}
\newcommand{\be}{\begin{equation}}
\newcommand{\ee}{\end{equation}}
\usepackage{makeidx}  

\usepackage{cite}
\usepackage{url}

\usepackage{amsfonts}
\usepackage{xspace}
\date{}

\title{Information-Theoretic Considerations Concerning the Origin of Life}
\author{Christoph Adami$^{1,2,3}$}

\begin{document}

\maketitle

\begin{flushleft}
$^{\bf{1}}$ Department of Microbiology and Molecular Genetics, Michigan State University, East Lansing, Michigan, USA.
\\
$^{\bf{2}}$ Department of Physics and Astronomy, Michigan State University, East Lansing, Michigan, USA.
\\
$^{\bf{3}}$ BEACON Center for the Study of Evolution in Action, Michigan State University, East Lansing, Michigan, USA.
\\
$^{\ast}$ E-mail: adami@msu.edu
\end{flushleft}

\begin{abstract}Research investigating the origins of life usually focuses on exploring possible life-bearing chemistries in the pre-biotic Earth, or else on synthetic approaches. Little work has been done exploring fundamental issues concerning the spontaneous emergence of life using only concepts (such as information and evolution) that are divorced from any particular chemistry. Here, I advocate studying the probability of spontaneous molecular self-replication as a function of the information contained in the replicator, and the environmental conditions that might enable this emergence. I show that (under certain simplifying assumptions) the probability to discover a self-replicator by chance depends exponentially on the rate of formation of the monomers. If the rate at which monomers are formed is somewhat similar to the rate at which they would occur in a self-replicating polymer, the likelihood to discover such a replicator by chance is increased by many orders of magnitude. I document such an increase in searches for a self-replicator within the digital life system avida.
\end{abstract}

Understanding and explaining the origin of life on Earth are perhaps the most difficult problem that science faces~\cite{Morowitz1992,Deamer1994,DeDuve1995,Koonin2011}. There are difficulties everywhere: most likely, there is no remnant of the original set of molecules that began their fight against the second law of thermodynamics (but see the discussion in~\cite{Benneretal2004,Daviesetal2009}), so that the molecules in use by the most primitive life extant in the biosphere today is almost completely different from those molecules that began it all~\cite{Joyceetal1987}. In view of all the problems that an RNA world scenario creates in terms of an ``impossible" chemistry~\cite{LevyMiller1998,Sutherland2010}, we cannot even assume that RNA (in its present form) was part of the ancient equation.

Given the unfathomable number of variables (geological, chemical, and environmental), is it even worth while pursuing an answer to a question we can barely formulate? Here I propose (as others have before me~\cite{Joyce2012,WalkerDavies2013}) that in parallel to the ongoing work exploring the conditions of the pre-biotic Earth and the possible chemistries that could have given rise to a chemical system supporting self-replication, we ought to pursue theoretical research that is disconnected from a particular chemistry; that instead focuses entirely on information-theoretic aspects of the problem, as well as simulations of (and with) artificial (abstract) chemistries. 

Information-theoretic considerations are, after all, the only thing we can be certain about in this wilderness of ideas and uncertainties concerning the emergence of a biosphere. In order to be considered living, a system must be able to maintain information on a time scale that exceeds (likely by many orders of magnitude) the abiotic scale (defined below)~\cite{Adami1998}. Indeed, if for a moment we define information as the difference between the maximum (thermodynamical) and the actual entropy of the system, then certainly an abiotic system will represent a vanishing amount of information unless it begins in a non-equilibrium state and approaches thermodynamic equilibrium. Let's agree to call this timescale--the time it takes for a non-equilibrium system to approach maximum entropy--the abiotic scale. Living systems can stay away from maximum entropy for much longer, indeed arbitrarily long (the biotic time scale is, for all we know, only limited by the existence of the biosphere). It is then this ability: to persist in a state of reduced entropy for biotic as opposed to abiotic time scales, that defines a set of molecules as living, and this set of molecules must achieve that feat via the self-replication of information. 

From this point of view, the transition from non-life to life has occurred if a thermodynamical system permanently moves from maximum entropy $H_{\rm max}$ to $H_{\rm max}-\Delta H$, where $\Delta H$ is the entropy deficit, or information. I argue here that how likely it is that such a transition happens spontaneously depends mostly on the size of $\Delta H$, and I will investigate here some quantities that are bound to affect the size of $\Delta H$ without referring to {\em any} aspects of the local chemistry that gives rise to that information. In this manner, I hope we can learn something about the environmental characteristics we should look for in candidates for origin-of-life scenarios, characteristics that hopefully go beyond those that we are already investigating. 

In the following, I will focus on potentially self-replicating molecular sequences of $L$ monomers. Obviously, this constrains potential origin-of-life scenarios, but not as much as one might think. First, for the information-theoretic treatment that follows, it does not matter whether the replicator is a single contiguous sequence or instead a set, where a number of sequences are replicating each other in an autocatalytic cycle~\cite{EigenSchuster1979} in such a manner that the end result is the self-replication of the information contained in the set. It is possible in principle to calculate the amount of information contained within the autocatalytic set (this is potentially non-trivial because of likely redundancies within the set of molecules), in which case the analysis that follows applies to this (non-redundant) information\footnote{Note that information is by its definition non-redundant (in the sense that the information contained in two identical sequences is equal to the information contained in one of them), but we can imagine that a set of molecules that all share some information is represented by one other sequence that encodes all of the set's information in a non-redundant manner.}. After all, when our DNA is replicated, this also happens via an intermediary of a myriad of other molecules, all of which are, however,  encoded within the chromosomal or mitochondrial DNA. 

Second, constraining our self-replicators to a limited number of possible monomers it could use (the ``repertoire") is only a weak constraint, because in an information-theoretic treatment of molecular chemistries we can ignore any monomers that are rarely used: they contribute only marginally to entropies (and hence information) on account of the $p_i\log p_i$ form of the Shannon entropy. Third, while it is easier to treat sequences of fixed length, the formalism can easily be extended to treat molecules of varying length. Focusing on linear polymers, however, is a restriction that ultimately might not turn out to be warranted. Indeed, it is possible to encode information in molecular assemblies that are not linear chains~\cite{Segreetal1998,Segreetal2000}, and for those my treatment would have to be modified. 

I first calculate the fraction of molecules of length $L$ (with monomers drawn from an alphabet or repertoire of size $D$) that are {\em functional}--in the sense that they display the ability to self-replicate. In the following, I do not bother about an error rate, because I simply assume that even if the self-replication is imprecise, there will be a finite fraction of molecules that will be replicated accurately. I assume this because were the error rate too high then information cannot be maintained, and indeed the sequence of symbols is strictly speaking not information, precisely {\em because} it cannot persist. Note that the critical error rate depends on the population size, because in very large populations a few error-free sequences might be generated even if most copies are flawed, which is all that is required for the maintenance of information. All these statements can be made precise, but I won't bother to undertake this here. 

Let us first define the fraction of molecules $F(x>0)$ with replication rate $x>0$ as
$F(x)=N_\nu/N$, where $N_\nu$ is the number of self-replicators (at any rate $x$ or larger), and $N$ is the total number of possible molecules of that length. If we assume (I will shortly relax this assumption) that all potential self-replicators appear with equal probability, then the Shannon entropy (also called ``uncertainty")  of the self-replicating ensemble is $H=\log N_\nu$. This uncertainty is related to the probability of finding such a self-replicating sequence by chance in a set of other sequences. The probability to find such a sequence among all {\em possible} sequences is of course just $F(x)$.
The information self-replicators have (about what it takes to self-replicate in their world) is the difference between the maximal and the actual entropy~\cite{AdamiCerf2000,Adami2004,Adami2012}
\be
I_S=H_{\rm max}-\log N_\nu =L-H\;, \label{info}
\ee
because the entropy of a random polymer of length $L$ is $H_{\rm max}=L$ mers (if we take logarithms with respect to base $D$)~\cite{Adami2004}. Because we also have $H_{\rm max}=\log N$, Eq.~(\ref{info}) implies a relationship between the fraction of functional molecules and their information content as suggested by Szostak~\cite{Szostak2003} $I_S=-\log F(x>0)$, but here I will go beyond this relationship and explore a more accurate estimate of $H$ that takes into account the composition of the polymer in terms of monomers, the relative frequency of the polymers, and the rate at which each monomer is being produced abiotically.

Let us first relax the assumption that all possible replicators appear with equal frequency in the population. Let us instead enumerate all replicators in terms of their genotype $i$. Then, the entropy of self-replicators is
\be
H=-\sum_{i=1}^{N_\nu}p_i\log_D p_i\;, \label{ent}
\ee
where $p_i$ is the likelihood to find genotype $i$ in an infinite population. Of course, we do not know those probabilities as infinite populations do not exist, so we will have to estimate them. We do this by writing the entropy of the molecules in terms of the entropy of monomers. Indeed, if all monomers in the sequence were independent of each other, then the entropy of the random variable that represents the $i$th polymer
\be
X_i=X^{(1)}_i\times X^{(2)}_i\times ... \times X^{(N)}_i\
\ee
could simply be written as the sum of the entropy of each monomer variable $X^{(j)}$ in that sequence
\be
H(X_i)=\sum_{j=1}^L H(X_i^{(j)})\;. \label{enti}
\ee
But we know from general arguments that monomers in a biological polymer are not independent, because if they were, self-replicators would be easy to find. Indeed, if monomers were independent, then the time it takes to find a particular sequence of length $L$ is of the order $L$ (while it is of the order $D^L$ in the worst-case scenario). However, information theory allows us to write the entropy (\ref{ent}) in terms of monomer entropies as well as correlation entropies. Correlation entropies between pairs of monomers are called ``information", but higher order correlations can exist too. If we define the Shannon information between two residues $i$ and $j$ as $H(i:j)$, the correlation entropy between three residues $i$, $j$, and $k$ as $H(i:j:k)$ and so on, we can write (\ref{ent}) as\footnote{This expression goes back to Fano's book, where he calculates the mutual information between an arbitrary number of events, see~\cite[p. 58]{Fano1961}.}
\be
H=\sum_{i=1}^L H(i)-\sum_{i>j}^LH(i:j)+\sum_{i>j>k}^LH(i:j:k)-\cdots  \label{full}
\ee
In (\ref{full}), the sum goes over alternating signs of correlation entropies, culminating with a term $(-1)^{L-1}H(1:2:3:\cdots :L)$.  

So now, the information contained in a molecule is
\be
I_S=L-H=L-\sum_{i=1}^L H(i)+\sum_{i>j}^LH(i:j)-\cdots\;.
\ee
Let me assume for a moment that we do not have to worry about the higher order informations $H(i:j)$, $H(i:j:k)$ and so on, in the sense that these terms will be smaller than all the first order terms $L-\sum_i H(i)$. This is not at all an obvious assumption, because each of the higher-order terms might be small, but there is an exponentially increasing number of them as the order increases. We can start worrying about these terms when everything else is said and done: at the moment let me just mention that I have seen very few cases where terms of the order $H(i:j:k)$ or higher play a role, while pairwise correlations such as $H(i:j)$ can play very important roles indeed~\cite{AdamiCerf2000,GuptaAdami2014}. 

For the sake of the argument, let me just consider the size of $I_S=L-\sum_i^LH(i)$. Such a term might be large, in particular if the positions are fairly well conserved ($H(i)\approx0$). By the preceding arguments, such a sequence will be very unlikely to emerge by chance, as this probability is
\be
P_0=F(x)=D^{-I_S}\;.
\ee
Now let me make one other simplifying assumption, namely that the entropy at each site is roughly the same, namely $H_b$: (`$b$' for ``biotic") 
\be
H(i)\approx H_b  \ \ \ \ \forall i\;,
\ee
Such an assumption is of course ludicrous when we think about the per-site entropy of known biomolecules, which varies tremendously from site to site (see, e.g.,~\cite{Adami2004,Adami2012,GuptaAdami2014}). Let us thus say that $H_b$ is really the average per-site entropy
\be
H_b=\frac1L\sum_i^LH(i)
\ee
so that
\be
I_S=L-LH_b=L(1-H_b)\;. \label{info0}
\ee
What sets the value of $H_b$? At each site, the entropy is determined by how often any particular monomer is found there on average in a typical functional protein. Suppose that each monomer appears on average with probability $q_j$ in such an informational molecule. The entropy of an average site is then
\be
H_b=-\sum_{j=1}^D q_j\log_D q_j\;.
\ee
If each monomer occurs roughly with the same frequency $q_j\approx 1/D$, then $H_b\approx1$, and $I_S=0$. Indeed, it is not possible to encode information in such a way, unless information is stored in higher order correlations. 
In Eq.~(\ref{info0}) we assumed that the set of all possible molecules had entropy $L$, which came from the assumption that in random (abiotic) molecules, each monomer did indeed appear with probability $1/D$. What if that was not the case? 

What if, by chance, monomers in random molecules have different frequencies? Indeed, amino acids in abiotic proteins do not at all occur at equal frequencies. Rather, their abundance is dictated by the rate at which they form abiotically (see, e.g., \cite{Dornetal2011} and references therein). Let's say this abundance distribution is
$\pi_j$, with entropy 
\be
H_a=-\sum_{j=1}^D\pi_j\log_D\pi_j\;,
\ee
and the information is then
\be
I_S=L(H_a-H_b)\;,
\ee
We now see that if by sheer luck the abiotic entropy $H_a$ is close to the biotic one $H_b$, then the entropy gap can be made arbitrarily small. And because the likelihood to find a sequence with information $I_S$ by chance is $D^{-I_S}$,
such a reduction in the entropy gap could affect the likelihood of spontaneous emergence of $I_S$ tremendously.

Imagine for example that the biotic distribution $q_j$ is fairly close to the abiotic one, that is $q_j=\pi_j(1+\veps_j)$,
where $\veps_j\ll1$ is symmetrically distributed around 0, so that $\sum_j\veps_j=0$. Then
\be
H_a-H_b=-\sum_j(\pi_j\log_D\pi_j-q_j\log_D q_j)\approx \sum_j\pi_j\veps_j^2 +{\mathcal O}(\veps^4)\;. \label{gap}
\ee
Note that because I assumed that the biotic distribution $q_j$ is derived from the abiotic one, the expression for the entropy gap (\ref{gap}) is guaranteed to be positive.

How big of a difference does a reduced entropy gap make? It can be dramatic, as I will now show. Consider for example the entropy gap for self-replicators in the digital-life system avida~\cite{AdamiBrown1994,Adami1998,WilkeAdami2002,OfriaWilke2004,Adami2006}, which is a system in which self-replicating computer programs mutate and evolve to adapt to a fitness landscape that can be specified by the user. A simple self-replicator can be written in avida that takes only 15 instructions taken from an alphabet of $D=26$ (these 15 mers are the equivalent of about 70 bits of information, as 15 $\times \log_2{26}\approx 70.5$). The probability to find such a self-replicator by chance is rather low:
\be
P_0=26^{-15}\approx 6\times 10^{-20}\;. \label{p0}
\ee 
If we would test a million random sequences a second on a parallel cluster of 1,000 CPUs, this search would take over 500 years. Now, it makes no sense to test sequences of length 15 because the per-site entropy of such a compressed replicator is about zero, so there is virtually no redundancy. Let us imagine that instead we test sequences of length 100 that have the same  information content of 15 mers. In that case, we expect the average per-site entropy of the self-replicator to be about 0.85 mers, so that $I_S=100\times(1-0.85)=15$, that is, the per-site entropy gap for self-replicators of this type is $\Delta H=0.15$. $P_0$ is of course still given by (\ref{p0}), which still all but rules out a random search. 

What if we change the abiotic entropy to the one that we expect for a replicator? We can estimate this biotic entropy $H_b$ from the entropy profile for avidians studied in~\cite{Dornetal2011}, reproduced in Fig.~1 for a typical case. 
\begin{figure}[htbp] 
   \centering
   \includegraphics[width=4in]{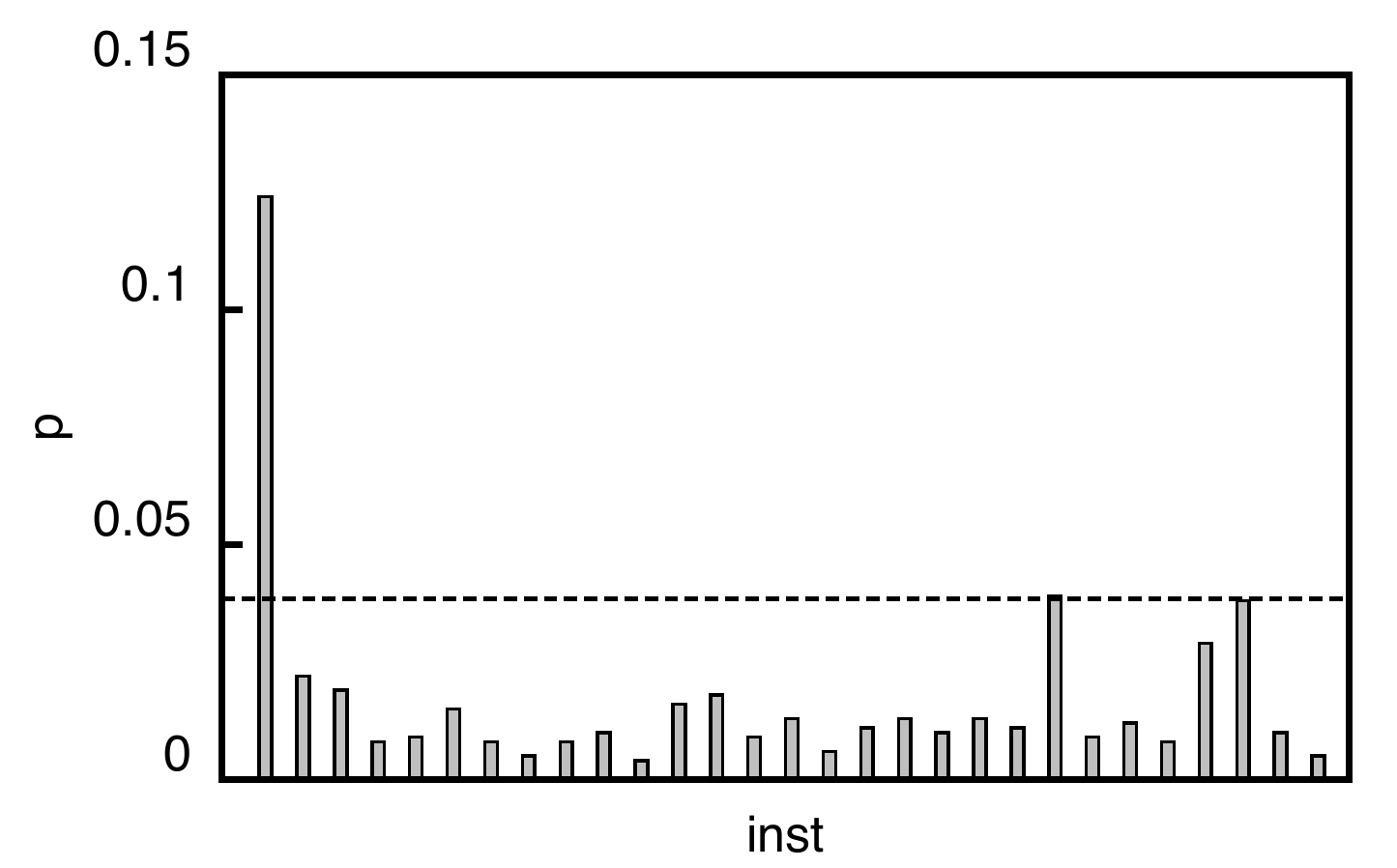} 
   \caption{Probability distribution of instructions (in random order on the $x$-axis) for an adapted avidian self-replicator with D=29 (from\protect\cite{Dornetal2011}). The largest fraction is the {\tt nop-A} instruction which is special in avida as it is used to initialize empty instructions. The dashed line is the uniform prior, with entropy $H_a=1$.}
   \label{fig:example}
\end{figure} 
This distribution has an average entropy of $\approx 0.87$ mers per site. Let us imagine then that avidian instructions are not produced with equal likelihood $H_a=1$ (the uniform prior given by the dashed line in Fig.~1) but with an entropy more like the actual observed entropy, say $H_a\approx 0.9$. This innocuous change will change the probability $P_0$ to the probability $P_\star$
\be
P_0\to P_\star = 26^{-100(0.9-0.85)}=26^{-5}\;.
\ee
The effective information appears to have been reduced from 15 to 5! The probability to find the self-replicator now is $P_\star\approx 8.8\times 10^{-8}$. When Rupp et al.\ performed a random search for self-replicators of length 100 using the {\em biased} $H_a$~\cite{Ruppetal2006}, that is, by replacing monomers not uniformly randomly but in a biased manner according to the probabilities we see in Fig.~1, they found 10 self-replicators in the $2\times 10^{8}$ they tried, that is, a rate of $5\times10^{-8}$ which is well in line with the reduced estimate above. But we should keep in mind that it is important here to not just mikmc the abiotic entropy (after all this can be achieved with any number of distributions), but to mimic the distribution that appears in biotic polymers, such as the one in Fig.~1 for avidians. 

Of course, the entropy gap for biomolecules could be much more severe than for digital organisms. We know, for example, that the information content of the HIV-1 protease (a 99 mer molecule) is approximately 75 mers~\cite{GuptaAdami2014}. The probability to find such a molecule by random search is astronomically low, $P_0=20^{-75}\approx 2.64\times 10^{-96}$. Now, of course the probabilities for the abiotic generation of amino acids is far from uniform. Indeed, if we restrict ourselves to only the 20 amino acids used in biochemistry, the abiotic distribution has a significantly smaller entropy than the biotic one (simply because the heavy amino acids are not formed abiotically at all, see~\cite{Dornetal2011}). It is clear that the biotic distribution has evolved to be far from the initial abiotic distribution, so we cannot anymore assume $q_j=\pi_j(1+\veps_j)$. 
Let us instead investigate what happens to $P_0$ if we change the abiotic distribution away from uniform.
In a 99-mer molecule with 75 mers of information, the average entropy per site is 24/99 $\approx 0.2424$, that is, we write the information content of the protease as $I_P=99*(1-0.2424)$. If we assume that the abiotic entropy per monomer is 0.5 rather than the 1.0 coming from the uniform assumption, the probability to find this molecule by chance changes to 
\be
P_\star=20^{-99(0.5-0.2424)}\approx 7.5\times 10^{-32}\;,
\ee
an enhancement of about 64 orders of magnitude. When such extreme amplification of likelihood of spontaneous discovery of information are possible, scenarios that were previously deemed impossible~\cite{Shapiro2000} could move much closer to reality.  For example (even though I do not believe that the first self-replicator was a protein) the abundance distribution of amino acids (and other molecules such as mono-carboxylic acids) found in meteorites is much closer to that found in sediment than the distribution found in synthesis experiments~\cite{Dornetal2011}. 

To be fair, nobody expects life to originate from a self-replicator of about 100 instructions drawn from an alphabet of 20-30. I presented these examples just to illustrate how abiotic monomer distributions that are close to the biotic distribution can enhance the likelihood of stumbling upon a rare self-replicator by chance by many orders of magnitude, just as the likelihood to find a self-replicator was enhanced tremendously in the biogenesis experiments with avida in Ref.~\cite{Ruppetal2006}. We should keep in mind, however, that self-replicating sets of molecules encoding about 86 bits of information have been constructed in the laboratory using existing RNA enzymes (even though only 24 of those bits were evolvable~\cite{LincolnJoyce2009}). The spontaneous emergence of this replicator is of course impossible from an unbiased library, as $P_0\approx 7.7\times 10^{-25}$, or 5 orders of magnitude smaller than the likelihood I calculated for the spontaneous emergence of an avidian self-replicator.

Clearly, more work is required to test the relationship between spontaneous emergence and biased monomer abundance distributions. We could, for example, study different prior (abiotic) distributions and conduct biogenesis experiments in avida just as in~\cite{Ruppetal2006}, to see if the correlation between reduced gap and enhanced emergence holds quantitatively. In the non-digital realm, we could repeat the experiment of Keefe and Szostak~\cite{KeefeSzostak2001}, who searched for proteins that bind ATP within a randomly library of $6\times 10^{12}$ randomly generated 80-mer proteins. Among this random set they found four proteins that bound ATP, suggesting that the information necessary for ATP binding is $I_s=-\log_D (2/3\times 10^{-12})\approx 9.4$ mers, a value not inconsistent with deletion experiments and sequence analysis. By creating a biased (rather than random) library that takes into account the amino acid frequency bias of the ATP binding proteins they found, it should be possible to increase the fraction of ATP binding proteins found by chance significantly. Indeed, Hackel et al.~ showed that biasing protein libraries with conserved domains (zero entropy regions) but also variable regions constrained by the entropic profile of functional molecules as described here, leads to an increased rate of finding functional proteins by chance~\cite{Hackeletal2010}, compared to the rate observed from unbiased libraries. 

Even though we still do not know which set of monomers gave rise to the first self-replicator (if ever there was one), the information-theoretic musings I have presented here should convince even the skeptics that, within an environment that produces monomers at relative ratios not too far from those found in a self-replicator, the probabilities can move very much in favor of spontaneous emergence of life. For every candidate chemistry then (where a self-replicator can be constructed),  we would should look for the environment that is best suited to produce it.

\subsubsection*{Acknowledgements}I thank Charles Ofria, Matt Rupp, Piet Hut, Jim Cleaves, and Tim Whitehead for discussions, and acknowledge the hospitality of the 2nd ELSI Symposium in Tokyo, where the ideas presented here were conceived. This work was supported by the National Science Foundation's BEACON Institute for the Study of Evolution in Action under contract No. DBI-0939454. 

\bibliography{OLEB}
\bibliographystyle{ieeetr}
\end{document}